\documentclass{winnower}

\usepackage{lmodern}
 
\usepackage[T1]{fontenc}

\begin{document}
\headheight=0pt
\headsep=0pt

\title{Symmetry-guided nonrigid registration: \\ the case for distortion correction in \\ multidimensional photoemission spectroscopy}

\author{R. Patrick Xian}
\author{Laurenz Rettig}
\author{Ralph Ernstorfer}
\affil{Department of Physical Chemistry, Fritz Haber Institute of the Max Planck Society,\authorcr Faradayweg 4-6, 14195 Berlin, Germany}

\date{}

\maketitle

\subsubsection*{\centering\abstractname{}}
Image symmetrization is an effective strategy to correct symmetry distortion in experimental data for which symmetry is essential in the subsequent analysis. In the process, a coordinate transform, the symmetrization transform, is required to undo the distortion. The transform may be determined by image registration (i.e. alignment) with symmetry constraints imposed in the registration target and in the iterative parameter tuning, which we call symmetry-guided registration. An example use case of image symmetrization is found in electronic band structure mapping by multidimensional photoemission spectroscopy, which employs a 3D time-of-flight detector to measure electrons sorted into the momentum ($k_x$, $k_y$) and energy ($E$) coordinates. In reality, imperfect instrument design, sample geometry and experimental settings cause distortion of the photoelectron trajectories and, therefore, the symmetry in the measured band structure, which hinders the full understanding and use of the volumetric datasets. We demonstrate that symmetry-guided registration can correct the symmetry distortion in the momentum-resolved photoemission patterns. Using proposed symmetry metrics, we show quantitatively that the iterative approach to symmetrization outperforms its non-iterative counterpart in the restored symmetry of the outcome while preserving the average shape of the photoemission pattern. Our approach is generalizable to distortion corrections in different types of symmetries and should also find applications in other experimental methods that produce images with similar features.


\section{Introduction}

\begin{figure}[htbp!]
	\begin{center}
		\includegraphics[scale=0.48]{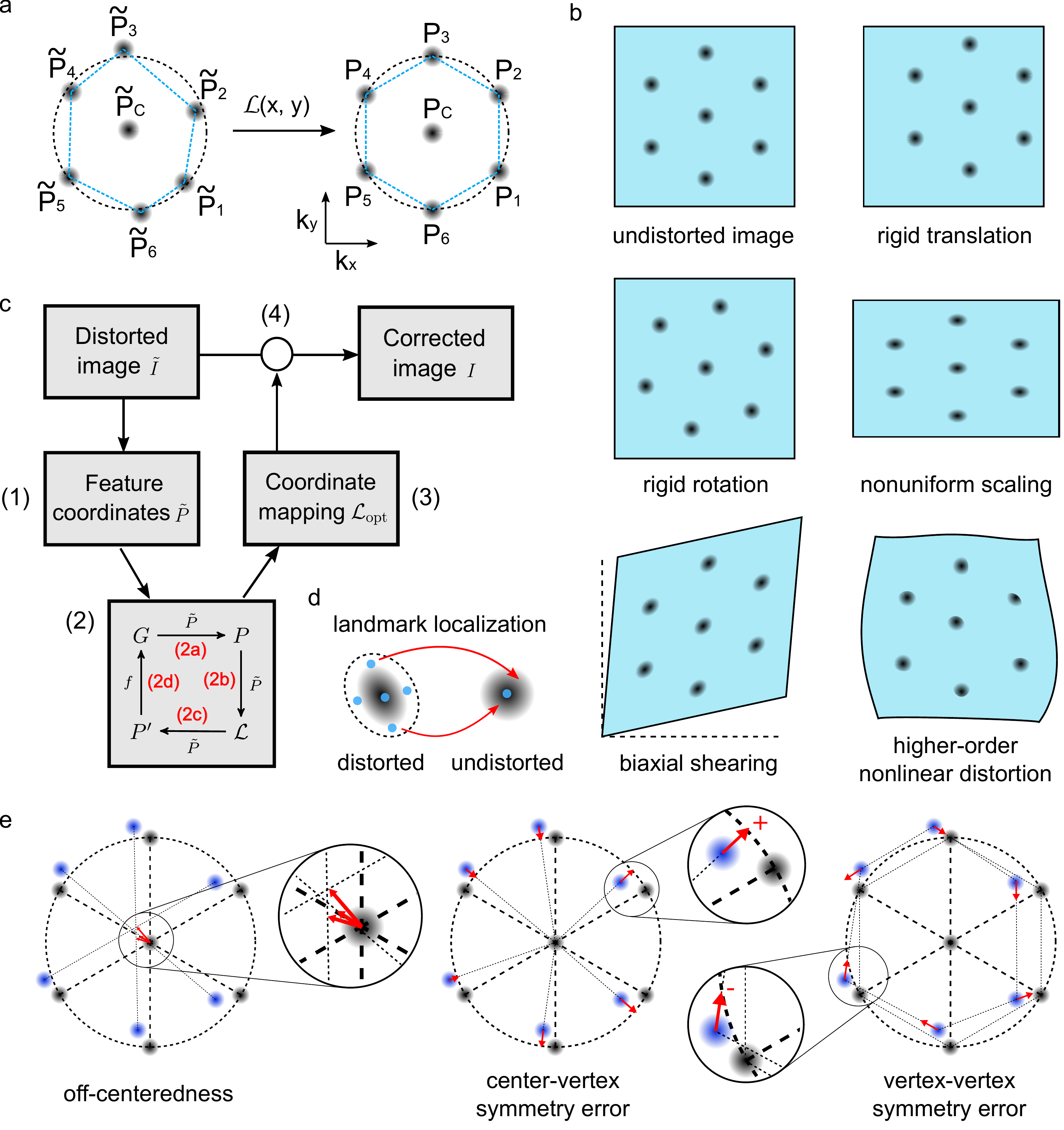}
	\end{center}
  \caption{(a) A cartoon example of distorted (left) and corrected (right) hexagonal pattern measured in photoemission spectroscopy. The dashed lines are for guiding the eye. (b) Comparison of undistorted image with those after rigid and nonrigid distortions. (c) Schematic of the proposed procedure for symmetry distortion correction. (1) Extraction of coordinates $\tilde{P}$ of the symmetry-related features from the distorted image $\tilde{I}$. The coordinates $\tilde{P}$ are used as inputs to the optimization loop in (2), which comprises of four parts: (2a) generation of target landmarks; (2b)-(2c) the landmark registration process; (2d) update of the generator $G$ under the symmetry constraints $f$. More details are given in the text. The loop in (2) yields (3), the optimized pixel coordinate transform $\mathcal{L}_{\text{opt}}$, which is then (4) applied to all pixels to correct the image distortion. Illustrations of (d) the origin of the landmark localization errors and (e) the geometrical representations of the error terms (Eq. \ref{eqn:fcenter}-\ref{eqn:fvv}) as red arrows in the optimization cost function. The distorted (blue) and undistorted (black) patterns share the center. The plus and minus signs next to the red arrows indicate whether the length should be added or subtracted, respectively.}
  \label{fig:loop}
\end{figure}

The use of symmetrization appears in various scientific contexts to assist data analysis. Generally speaking, symmetrizing an object requires a coordinate transform that enhances its symmetry with minimal alteration of the shape (e.g. curvature, area or volume) \cite{Mitra2007}. In solid state physics, the symmetry of the electronic band structure (EBS) of materials is often revealed in photoemission measurements with resolution of the photoelectron momenta \cite{Hufner2003,Damascelli2003}. Thanks to recent technological advances, characterization of the EBS in momentum $k_x$, $k_y$, and energy $E$ coordinates can now be achieved in \textit{en bloc} measurements using time-of-flight electron detectors, a technique named multidimensional photoemission spectroscopy (MPES) \cite{Schonhense2015,Schonhense2018}. The processing of MPES data often utilizes their symmetry for signal averaging, and for extracting and comparing the underlying physical quantities, such as the transition matrix elements \cite{Damascelli2003,Moser2017} and the electron self-energy \cite{Damascelli2003}, or to reconstruct the electronic orbitals of solid state materials \cite{Puschnig2009,Offenbacher2015}. However, systematic deviations in experimental data from the symmetry expected for the EBS, called symmetry distortion (see Fig. \ref{fig:loop}a-b), are often present due to various sources of imperfections in the instrument design and experimental setup, such as mechanical inaccuracies, incompletely shielded stray magnetic fields, sample geometry and surface quality, experimental conditions, including the lens alignment and voltage settings, the photon beam shape, etc. While some of the hardware imperfections may be calibrated using samples with a well-known band structure, the complication of the numerous factors varying by the sample and by the experiment renders a complete calibration of symmetry distortion impossible. Nevertheless, since these nonidealities result in perturbations to the trajectories of the photoemitted low-energy electrons, and subsequently, the deviation from perfect symmetry in the detected photoemission pattern, a post-processing algorithm to correct (see Fig. \ref{fig:loop}a) and quantify imperfect symmetries using coordinate transforms \cite{Glasbey1998,Holden2008} and image features is of great practical use. In established angle-resolved photoemission experiments measuring the 2D $k-E$ dependence using hemispherical analyzers, rescaling of the image coordinates may be used to minimize the symmetry distortion. For the distortion correction of 3D EBS mapping data measured by time-of-flight detectors or the deflector mode of the hemispherical analyzers, a symmetrization transform needs to be applied to at least the two momentum dimensions, $k_x$ and $k_y$, simultaneously. We describe an algorithm that achieves this.

Several symmetrization methods have been previously developed in different scientific contexts. For rotationally symmetric scenarios, a circularization procedure using directional rescaling \cite{Guichard2013} was developed for cryoelectron tomography. Alternatively, the basis expansion approach such as decomposition into Fourier series \cite{Gascooke2017} or spherical harmonics \cite{Greber2001} can also be used to symmetrize image patterns. In the studies on translation-symmetric systems, the strain field mapping technique used in high-resolution transmission electron microscopy \cite{Galindo2007} to characterize the local atomic misarrangements, and the stage drift correction algorithms \cite{Savitzky2018,Wang2018} may also be regarded as solving symmetrization problems. However, these approaches have limited applicability to MPES data due to the differences in their image features. Typical MPES data possess discrete rotational symmetry with anisotropic intensity modulations resulting from the matrix element effect \cite{Damascelli2003,Moser2017}, which should be considered in the algorithm design and adaptation for the purpose of image symmetrization.

Despite the complex and often unknown origins, the distortion in the photoemission pattern may be classified by visual inspection (see examples in Fig. \ref{fig:loop}a-b). We recognize the major types of distortions as (1) biaxial shearing and (2) nonuniform scaling, along with (3) higher-order nonlinear distortion (including perspective distortion and off-centeredness), which contribute in less amounts. Undistorting the image generally requires changing the distances between points by applying a nonrigid coordinate transform, which may be determined by matching the corresponding features in the distorted image (reference) and the undistorted one (target), similar to the technique called nonrigid image registration (i.e. alignment) \cite{Holden2008,Goshtasby2005}. However, in image symmetrization, the target image doesn't exist, so the problem is constrained only on the reference image side, whereas image registration is constrained on both the reference and the target sides. Previously, tuning of a single shearing parameter using point feature correspondences has been shown to improve the symmetry of the photoemission pattern \cite{Schonhense2018}, but its simplicity doesn't always guarantee sufficient symmetrization, hampering subsequent detailed data analysis. Our approach, the symmetry-guided nonrigid registration, rests on the observations that the outcome of symmetrization using image registration is affected by (1) the degrees of freedom in the coordinate transform of choice and (2) the quality of the feature correspondences such as the uncertainties in their exact positions, also known as the localization errors (LE, see Fig. \ref{fig:loop}d) \cite{Rohr2001,Rohr2001a}. To improve the former aspect, we resort to nonrigid transforms \cite{Glasbey1998,Holden2008,Goshtasby2005} that account for more types of distortions (see Fig. \ref{fig:loop}b). For the latter, we use an iterative framework to explore more spatial configurations of the feature correspondences to compensate for the LE and optimize for more symmetry in the outcome.

In the following, we first present the definition and characterization of the iterative symmetrization framework, followed by the mathematical description of its components in an example case of a polygon-type pattern with discrete rotational symmetry. The description of alternative cases are presented in the Supplementary Material section S1. We then compare the iterative and the non-iterative approaches using experimental MPES data and quantify their performances using our proposed symmetry metrics independent of the symmetrization algorithm. In addition, we provide an open-source software package, \textsf{symmetrize} \cite{symmetrize}, written in Python that implements the procedures described in this work for public use.

\section{Methods}
Symmetrization of MPES data makes use of the perceived symmetry of the photoemission pattern in the $k_x-k_y$ plane (see Fig. \ref{fig:loop}a). The pattern's geometric shape relates to the symmetry of the bulk or surface Brillouin zone of the material under study, and is generally more robust than the intensity symmetry against modulations by the matrix element effect in the photoemission process \cite{Note} and the experimental configuration, which, in combination, create complex anisotropic modifications to the measured intensities in the momentum space \cite{Moser2017}. The geometric pattern is defined by point landmarks. Those that are relevant for the perceived symmetry may be used for image registration based on landmarks (i.e. landmark registration) \cite{Goshtasby2005}. Typically, landmark registration employs a large amount of point correspondences between the reference and the target images obtained by a feature selector. The correspondences are putative and often have varied accuracies in their locations and relations such that a robust fitting algorithm is required to determine the transform parameters \cite{Szeliski2011}. For symmetrization, with only constraints on the reference side, it's possible to produce point correspondences using symmetry relations and a target (feature) generator with high consistency (i.e. genuine instead of putative correspondence). However, the position inaccuracies in the landmarks selection, also known as the landmark localization errors (LLE, see Fig. \ref{fig:loop}d) \cite{Rohr2001,Rohr2001a}, affects the symmetrization more severely with a small amount of correspondences, which is characteristic of MPES data symmetrization. Since LLE only exists in the reference for image symmetrization, we can compensate for it by relaxing the symmetry in the generated target landmarks. In practice, we iteratively optimize the parameters of the target generator to tune the amount of symmetry relaxation in order to improve the symmetrization outcome. The transform determined with the iterative approach improves on the result from using a fixed set of symmetric target landmarks. On the other hand, the optimization allows the use of constraints designed to simultaneously minimize the size change of the pattern, which is important for maintaining the image resolution (or electron momentum resolution in the case of MPES data). Moreover, the iterative approach is applicable to various types of coordinate transforms. We formulate the iterative symmetrization procedure as follows,

{\textit{Given a set of symmetry-related point landmarks $\tilde{P}$ sampled from the distorted reference image $\tilde{I}$, find a target point set $P = G(\alpha ; \tilde{P})$, related to $\tilde{P}$ by the generator function $G$ with parameters $\alpha$. $\tilde{P}$ and $P$ together determines a parametrized transform $\mathcal{L}$, such that the transformed point set, $P' = \mathcal{L}\tilde{P}$, satisfies the constraints $f$.}}

The iterative symmetrization is at the center of the overall distortion correction procedure, shown schematically in Fig. \ref{fig:loop}c. In the present work, we demonstrate the effectiveness of the iterative approach using two types of transform functions as $\mathcal{L}$. The first is the eight-parameter 2D projective transform (also known as a perspective transform or homography \cite{Szeliski2011,Hartley2004}), which corrects most of the biaxial shearing and nonuniform scaling. For a more complete distortion correction, we use the thin plate spline (TPS) \cite{Bookstein1989,Pincus2007} transform. In addition to these, other similar types of parametric transforms are also feasible \cite{Glasbey1998,Holden2008}. The constraints $f$ are formulated using geometric relations as a cost function for optimization. The point features used here may be generalized to other salient symmetry-equivalent image features such as lines, curves, etc.

The procedure in Fig. \ref{fig:loop}c starts with (1) a point set extracted using a landmark detection algorithm (e.g. 2D peak detection) or manual selection, it is ordered clockwise or counterclockwise using the opening angle between a fixed axis unit vector (e.g. along the image axis $x$ or $y$) and the vector connecting each point and the center or centroid. The ordered point set represents the vertices of a polygon, and we adopt the term vertex and center from now on. For a pattern without a clearly defined center, the centroid position of the vertices may be used. We denote the ordered reference and target point sets as $\tilde{P}$ and $P$, respectively, as shown in the example in Fig. \ref{fig:loop}a. The intermediate point set $P'$ relates to $P$ exactly through the transform $\mathcal{L}$. Note that $P = \{P_i\}, \, \tilde{P} = \{\tilde{P_i}\}, \, P' = \{P'_i\}$, excluding their respective centers $\tilde{P}_C$, $P'_C$ and $P_C$. All point positions are expressed in the homogeneous coordinate system \cite{Szeliski2011} for convenience of calculation, therefore, $\tilde{P}$, $P$, $P' \in \mathbb{R}^3$, with $P_i = (x_i, y_i, 1)^T, \, \tilde{P}_i = (\tilde{x}_i, \tilde{y}_i, 1)^T, \, P_i' = (x_i', y_i', 1)^T \,\, (i = 1, 2, ...)$. The first two dimensions in the homogeneous coordinates are the 2D Cartesian coordinates. For distortion correction using the projective transform, these point sets are related to one another in the optimization loop (see Fig. \ref{fig:loop}c) in the following ways.
\begin{alignat}{2}
(\textbf{2a}) \quad &P_i &&= G_i(\alpha_i;\tilde{P_i}) = s_iR(\theta_i)\tilde{P_i} \label{eqn:scalerot}\\
(\textbf{2b}) \quad &\mathcal{L} &&= DLT(\tilde{P}, P) \label{eqn:dlt}\\
(\textbf{2c}) \quad &P_i^\prime &&= \mathcal{L} \tilde{P}_i
\end{alignat}
In Eq. \ref{eqn:scalerot}, $P_i$ and $\tilde{P}_i$ are related by a scaling factor $s_i$ and a 2D rotation matrix $R(\theta_i)$. The collection of scaling and rotations constitutes the generator $G$ in Fig. \ref{fig:loop}c, with ${s_i, \theta_i}$ being the set of parameters $\alpha$. The transform $\mathcal{L}$ (in case of projective transform) is determined by the established method of direct linear transformation (DLT) \cite{Hartley2004}, and the single-pass solution $P_i'$ is the current best approximate to $P_i$, as expressed in Eq. \ref{eqn:dlt}. The labels (2a)-(2c) before the equations refer to the corresponding steps in Fig. \ref{fig:loop}c. The subsequent step (2d) refers to updating the parameters $\{s_i, \theta_i\}$ in the generator $G$ under the constraints or cost function $f$. The steps (2a)-(2d) lead to the sequential update $(G \rightarrow P \rightarrow \mathcal{L} \rightarrow P' \rightarrow G)$ of the quantities on the four corners in the optimization loop in Fig. \ref{fig:loop}c at every iteration.

The cost function $f$ in Fig. \ref{fig:loop}c is formulated using the sum of squared errors between the actual positions registered to ($P'$) and the ideal positions represented by the average distances as expected for a symmetric pattern, which will be discussed next. For a pattern with discrete even-order rotational symmetry (e.g. hexagon), it contains three error terms. The off-centeredness $f_{\text{centeredness}}$ is calculated by the distance between the actual center ($\tilde{P}_C$, if present), usually the $\Gamma$ point in the band structure mapping data, and the center determined by the vertices. The symmetry error comprises of two parts, one for the center-vertex symmetry $f_{\text{cvsym}}$, calculated using all vertices, and the other for the vertex-vertex symmetry $f_{\text{vvsym}}$, calculated using the nearest-neighboring vertices. The mathematical expressions of the geometric errors are given in the following for the case of a regular (i.e. equiangular and equilateral) polygon pattern with an even number of vertices (even-order rotational symmetry),
\begin{align}
f_{\text{centeredness}} &= \frac{2}{n} \sum_{i=1}^{n/2} \left\Vert \frac{P'_i + P'_{i + \frac{n}{2}}}{2} - \tilde{P}_C \right\Vert^2 \label{eqn:fcenter}\\
f_{\text{cvsym}} &= \frac{1}{n} \sum_{i=1}^n \left| \Vert P'_i - P'_C \Vert - \overline{\Vert\tilde{P}_i - \tilde{P}_C\Vert}\right|^2 \label{eqn:fcv}\\
f_{\text{vvsym}} &= \frac{1}{n} \sum_{\text{NN}} \left| \Vert P'_i - P'_j\Vert - \overline{\Vert\tilde{P}_i - \tilde{P}_j\Vert}_{\text{NN}}\right|^2 \label{eqn:fvv}
\end{align}
In Eq. \ref{eqn:fcenter}-\ref{eqn:fvv}, $n$ denotes the order of the rotation symmetry, e.g. $n=6$ for a hexagonal pattern. We use $\Vert \cdot \Vert$ to denote the Euclidean norm and $|\cdot|$ the absolute value of a scalar. The term $\overline{\Vert\tilde{P}_i - \tilde{P}_C\Vert}$ represents the average center-vertex distance, and $\overline{\Vert\tilde{P}_i - \tilde{P}_j\Vert}_{\text{NN}}$ the average nearest-neighbor (NN) vertex-vertex distance, both calculated in the coordinate system of the distorted image. For the hexagonal pattern, the average NN vertex-vertex distance is calculated from an average of $\Vert\tilde{P}_1-\tilde{P}_2\Vert$, $\Vert\tilde{P}_2-\tilde{P}_3\Vert$, $\Vert\tilde{P}_3-\tilde{P}_4\Vert$, $\Vert\tilde{P}_4-\tilde{P}_5\Vert$, $\Vert\tilde{P}_5-\tilde{P}_6\Vert$ and $\Vert\tilde{P}_6-\tilde{P}_1\Vert$. The formulation using average distances penalizes simultaneously the deviations of both the symmetry and the size change of the pattern. Extensions of the geometric errors based on Eq. \ref{eqn:fcenter}-\ref{eqn:fvv} to other scenarios are discussed in the Supplementary Material (see Section S1). The overall cost function $f$ is a numerical combination of all three terms described above,
\begin{equation}
f = f(P', \tilde{P}) = f_{\text{centeredness}} + f_{\text{cvsym}} + f_{\text{vvsym}}
\label{eqn:fsum}
\end{equation}
As the amount of distortion can vary with experimental conditions, one may introduce a weight to each contributor of the multi-objective cost function in Eq. \ref{eqn:fsum} in order to balance the contributors and steer the optimization process. The weighted cost function is, therefore,
\begin{equation}
f_{\text{w}} = w_{\text{c}} f_{\text{centeredness}} + w_{\text{cv}} f_{\text{cvsym}} + w_{\text{vv}} f_{\text{vvsym}}
\end{equation}
The weights may be estimated by the amount of distortion in a particular term. To solve for the optimal target point set $P_{\text{opt}}$ that compensates for the LLE in the reference image, we minimize $f$ (or $f_{\text{w}}$) with respect to the parameters $\{s_i, \theta_i\}$ in the generator $G$,
\begin{equation}
\{s_i^{\text{opt}}, \theta_i^{\text{opt}}\} = \underset{\{s_i, \theta_i\}}{\operatorname{argmin}} f(P',\tilde{P})
\end{equation}
Then, $P_{\text{opt}}$ can be determined using $\{s_i^{\text{opt}}, \theta_i^{\text{opt}}\}$ and Eq. \ref{eqn:scalerot}. It's not a strictly rotation-symmetric point set (see red crosses in Fig. \ref{fig:figcomparison}b, \ref{fig:figcomparison}d) and the deviation accounts for the LLE from the selection of $\tilde{P}$. The steps (3) and (4) in Fig. \ref{fig:loop}c correspond to determining the transform $\mathcal{L}_{\text{opt}}$ using Eq. \ref{eqn:dlt} and applying it to all pixels in the image, respectively. Likewise, the transform may also be applied to the single photoelectron event data directly recorded by the time-of-flight 3D detector. Extension of the iterative symmetrization approach to more complex features only requires to update the geometric constraints in Eq. \ref{eqn:fcenter}-\ref{eqn:fvv} using feature-specific metrics \cite{Goshtasby2005,Krig2016}. In the Supplementary Material, we have provided a walk-through (see section S2) of the steps shown in Fig. \ref{fig:loop}c using the relevant functions of the \textsf{symmetrize} software package.

To quantify the level of symmetry within the image using point positions, we propose to use metrics different from the error terms in Eq. \ref{eqn:fcenter}-\ref{eqn:fvv}. In this way, the metrics will also allow for comparison with future iterations of similar algorithms. In this context, the continuous symmetry measure (CSM) \cite{Zabrodsky1993} adapted to rotationally symmetric objects \cite{Zabrodsky1992,Frey2007} serves the purpose. The measure is invariant under similarity transformation (translation, rotation and uniform scaling) \cite{Szeliski2011} and bounded within the range $\left[0, 1\right]$, with $0$ representing complete symmetry, $1$ for complete asymmetry, and the values in between for various degrees of distorted symmetry. In addition, we introduce the area retainment measure (ARM), $\tanh\left|1 - \frac{A}{A_0}\right|$, to quantify the area change before and after the distortion correction. $A_0$ and $A$ represent the areas of the polygonal region of interest in the uncorrected and corrected images, respectively. The hyperbolic tangent function is used to squash an unbounded function in $\left[0, +\infty\right)$ down to a bounded function in $\left[0, 1\right]$. Similar to CSM, the closer the ARM is to zero, the less the area change (or the higher the area retainment), and the closer it is to one, the more area change. We define the symmetry recovery score (SRS) $S$ in the outcome as the mean of the CSM and the ARM,
\begin{equation}
S = \frac{1}{2} \left(CSM + \tanh\left|1 - \frac{A}{A_0}\right|\right)
\end{equation}
The SRS reflects the balance between the level of symmetry and the amount of shape perturbation in the symmetrization process \cite{Mitra2007}. Further details on the computation of the proposed symmetry scores are given in the Supplementary Material (see section S3).

\section{Results and discussion}
\begin{figure}[p!]
  \centering
  \begin{minipage}[c]{\textwidth}
  \centering
    \includegraphics[scale=0.48]{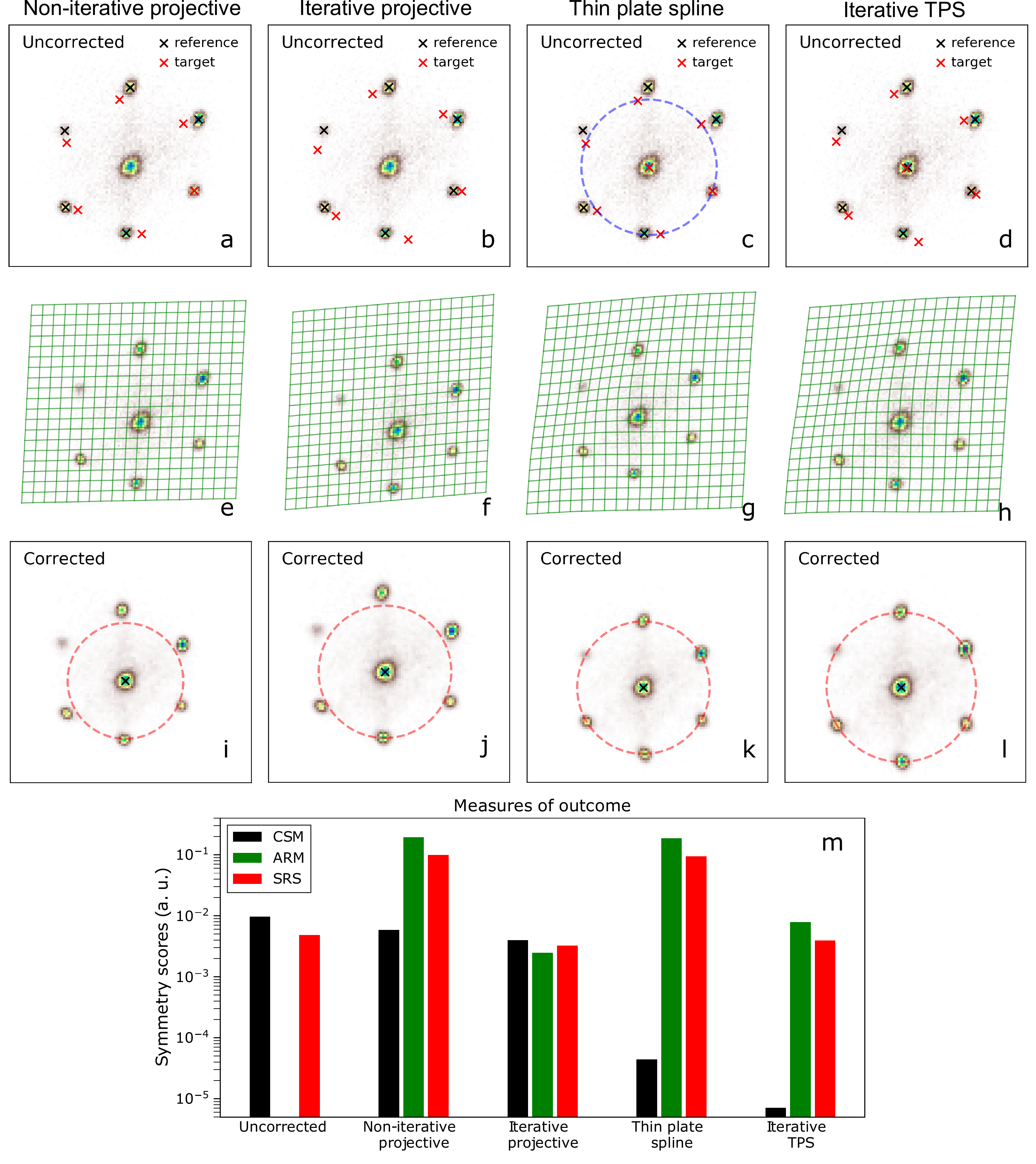}
    \caption{Comparison of the non-iterative and the iterative approaches of symmetrization using an energy slice from the WSe$_2$ band structure mapping measurement. (a)-(d) show the reference (black cross) and target (red cross) points in the uncorrected image used for each method shown in the title. (e)-(h) show the deformed grid (or deformation field) obtained from the corresponding symmetrization transform. (i)-(l) show the corrected images after applying the transform to all pixels in each case. An additional rotation is applied afterwards to align a symmetry axis with an image axis for presentation. The dashed blue and red circles in (c) and (i)-(l), respectively, guide the eye for symmetry evaluation. (m) The continuous symmetry measure (CSM), the area retainment measure (ARM) and the symmetry recovery measure (SRS) computed for the uncorrected and corrected images for quantitative comparison (see values in Table \ref{tab:symscores}).}
    \label{fig:figcomparison}
  \end{minipage}
\end{figure}

\begin{figure}[ht!]
  \begin{center}
    \includegraphics[scale=0.5]{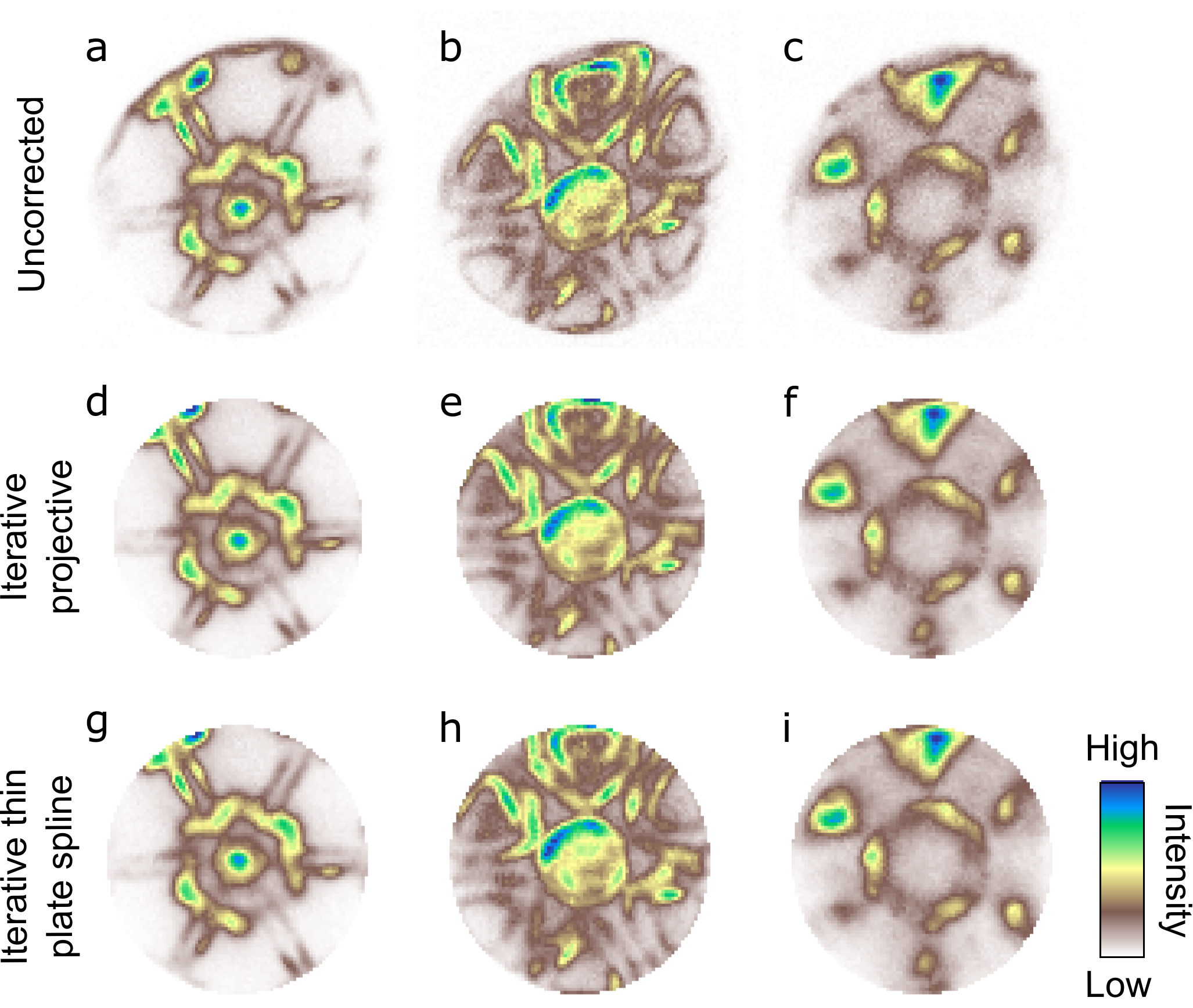}
    \caption{Application of the symmetrization transforms obtained from the iterative approaches shown in Fig. \ref{fig:figcomparison} to image slices at other photoelectron energies in the lower valence bands of WSe$_2$ from the same measured dataset. (a)-(c) show the uncorrected images, (d)-(f) show the corrected images using the projective transform and (g)-(i) using the thin plate spline, both parametrized from iterative estimation. The circular masks are used in (d)-(i) to remove the extraneous intensity features near the aperture edge.}
    \label{fig:figothers}
  \end{center}
\end{figure}

The band structure mapping measurement was carried out using a commercial electron momentum microscope (METIS 1000, SPECS GmbH) and a home-built table-top high harmonic generation-based pulsed extreme UV photon source at 21.7 eV \cite{Puppin2018}. Single crystalline tungsten diselenide (WSe$_2$) samples, which possess a sixfold rotational symmetry in its Brillouin zone, were used for the meausurements. The crystals were purchased from HQ Graphene and downsized and glued to a sample holder using conductive epoxy before measurements. Upon transferring to the measurement location, the sample was cleaved in vacuum using a cleaving pin. The photoelectron events recorded by the momentum microscope were converted to volumetric data via multidimensional histogramming in post-processing \cite{Xian}. Energy slices from the uncorrected data are shown in Fig. \ref{fig:figcomparison}a-d and Fig. \ref{fig:figothers}a-c. The horizontal and the vertical axes shown in each image are the photoelectron momenta $k_x$ and $k_y$, respectively.
\begin{table}[htbp!]
  \begin{center}
    \begin{tabular}{ c|c|c|c }
      \hline
      Methods & CSM & ARM & SRS \\
      \hline
      Uncorrected & $9.69\times10^{-3}$ & $0$ & $4.85\times10^{-3}$ \\ 
      Non-iterative projective & $5.84\times10^{-3}$ & $1.94\times10^{-1}$ & $9.98\times10^{-2}$ \\ 
      Iterative projective & $3.99\times10^{-3}$ & $2.47\times10^{-3}$ & $3.23\times10^{-3}$ \\
      Thin plate spline & $4.44\times10^{-5}$ & $1.88\times10^{-1}$ & $9.42\times10^{-2}$ \\
      Iterative TPS & $7.16\times10^{-6}$ & $7.88\times10^{-3}$ & $3.98\times10^{-3}$ \\      
      \hline
    \end{tabular}
  \caption{Symmetry scores before and after distortion \\ correction of the image with hexagonal pattern.}
  \label{tab:symscores}
  \end{center}
\end{table}

Distortion correction was first carried out using an energy slice close to the valence band maximum of the WSe$_2$ band structure mapping data, because its sharp features, the $K$, $K'$ and $\Gamma$ points \cite{Straub1996,Riley2014}, allow relatively precise landmark localization. The outcomes using both the non-iterative and the iterative methods are shown in Fig. \ref{fig:figcomparison}a-l, including the energy slice before and after the correction, the landmarks and the deformation field retrieved from the nonrigid registration.

The symmetry scores calculated for the outcomes are presented in Table \ref{tab:symscores} and visualized in Fig. \ref{fig:figcomparison}m. In both distortion correction using projective and TPS transforms, the iterative approach shows quantitative improvements from its non-iterative counterpart by all metrics, which demonstrates the effectiveness of the constraints in enhancing the level of rotational symmetry while preserving the overall shape of the region defined by the landmarks. The center point was not used in the parameter estimation for the projective transform in order to avoid overconstraining the optimization, because of its relatively fewer number of parameters. The order of magnitude differences between the outcomes using the projective and the TPS transforms is due to the difference in the degrees of freedom these transforms possess.

Next, we applied the same symmetrization transform for the hexagon pattern to other energy slices with distorted symmetry from the same volumetric measurement (see examples in Fig. \ref{fig:figothers}). Even though some of the images have less obvious symmetry-equivalent landmarks, the transformed images show well-restored symmetry. In other measurements without strictly point-like symmetry equivalents, one may use manually selected approximate landmarks as inputs to the algorithm to solve for the symmetrization transform. The procedures described in the Methods section were implemented in the Python package \textsf{symmetrize} \cite{symmetrize} and tested on a number of experimental datasets. For specific alternative use cases, the package may be extended by adding user-defined cost functions.

\section{Conclusions}
We have formulated image symmetrization in terms of nonrigid image registration guided by symmetry constraints and have proposed an iterative framework to compensate for the localization errors of point landmarks in the reference image. The framework incorporates the knowledge of the pattern's symmetry to construct the image registration target and to formulate the geometric constraints for iterative optimization. Using distorted electronic band structure mapping data as a test case, we compared the iterative and non-iterative approaches for estimating the symmetrization transform for a two-dimensional image with distorted discrete rotation-symmetric elements. By the use of scalar metrics to quantify the symmetry and the area change, we show that the symmetry constraints in the iterative approach allow further reduction of symmetry distortion while better preserving the area of the pattern, thereby outperforming its non-iterative counterpart. In addition to multidimensional photoemission spectroscopy, our approach should find convenient adoption and adaptation in other experimental methods in which distortion correction is necessary to recover the symmetry in the measured data to improve their analyses. The open source codebase enables further development and reuse by a broader community. Finally, the symmetrization procedure is an important step for the development of a data processing pipeline to extract symmetry-related quantities and physical insights from multidimensional photoemission measurements.

\section*{Acknowledgements}
This project has received funding from the Max Planck Society and from the European Research Council (ERC) under the European Union's Horizon 2020 research and innovation program (Grant Agreements No. ERC-2015-CoG-682843). We thank S. Dong, Y. W. Windsor, M. Dendzik, M. Puppin and C. W. Nicholson for contributions in the instrument construction and S. Dong for performing the measurement. We thank M. Wolf for his support, P. Hofmann for helpful discussion, S. Sch{\"u}lke and G. Schnapka for providing the computing infrastructure. L. Rettig acknowledges funding from the DFG in the Emmy Noether program under grant number RE 3977/1.


\bibliographystyle{elsarticle-num}
\bibliography{distortioncorr}

\begin{thebibliography}{10}
\bibitem{Note}
{Note that the matrix element effect alone can already cause the pattern to
  look asymmetric in its intensity distribution, even when no other artifact is
  present.}
\end{thebibliography}

\end{document}